# FACING THE FACTS

Patrick O'Beirne
Systems Modelling Ltd, Gorey, Co. Wexford, Ireland
pob <at> sysmod.com, www.sysmod.com

## *Abstract*

*Human error research on overconfidence supports the benefits of early visibility of defects and disciplined development. If risk to the enterprise is to be reduced, individuals need to become aware of the reality of the quality of their work. Several cycles of inspection and defect removal are inevitable. Software Quality Management measurements of defect density and removal efficiency are applicable. Research of actual spreadsheet error rates shows data consistent with other software depending on the extent to which the work product was reviewed before inspection. The paper argues that the payback for an investment in early review time is justified by the saving in project delay and expensive errors in use.*

*'If debugging is the process of removing bugs, then programming must be the process of putting them in' - Anon.*

## Initial questions (answers later)

1) Are defects good or bad?
2) If your reviewer tells that they found no defects in a spreadsheet, is that good or bad?

## Introductory definitions

To err is human; people make errors. In his book *Human Error* [3], James Reason uses 'error' as a generic term that can be categorised as slips in execution (eg typos) or mistakes in judgment (eg using the wrong word). In programming, human error creates defects (aka faults) in the software which can be immediately visible or latent. Depending on the conditions in production use, some defects may give rise to possibly many incidents of failure, and some may never be encountered. These incidents are reported by the user as fault reports, often called bug reports.

Defect Density is a standard quality measure of the number of defects per thousand lines of code (KLOC) or Function Points (FP) or, in the spreadsheet domain, Unique Formulas (UF). The defect density of a product follows a cumulative curve which rises at the start as defects are created and falls later as they are found and removed.

Injected Defects are defects put into the product usually due to mistakes which people make; but sometimes software is intentionally seeded with known defects in order to measure the efficiency of the review process. The Defect Injection Rate (DIR) is the percentage of work that is defective. You never know how many injected defects you have in your product – you can only estimate them from previous experience or benchmarks.

Removed Defects are defects which are identified by processes such as design review, code inspection, execution test, or (most expensive of all) user experience. They are also usually classified by severity in order to prioritise the fixing effort and then removed by rework. The Defect Removal Efficiency (DRE) is the percentage of defects detected by the find and fix processes.

The number of defects detected/fixed per unit of time (or effort) is useful for estimating when a product is ready for release.

## Reported Error Rates in Large Spreadsheets

Robert Lawrence [1] reported these statistics on the thirty most financially significant projects that Mercer Finance & Risk Consulting reviewed year-ending June 2004:

- Average 2,182 unique formulae per model
- Average 151 issues raised during the initial review
- Average six revisions to produce a model that could be signed-off
- One spreadsheet needed 17 revisions to resolve 239 issues
- Average 7% defect injection rate, 75% defect removal efficiency

Figure 1 shows that the defect rate for a model can be estimated as 62 plus 0.41 times the number of unique formulas (UF); or, 2.6 times the square root of UF.

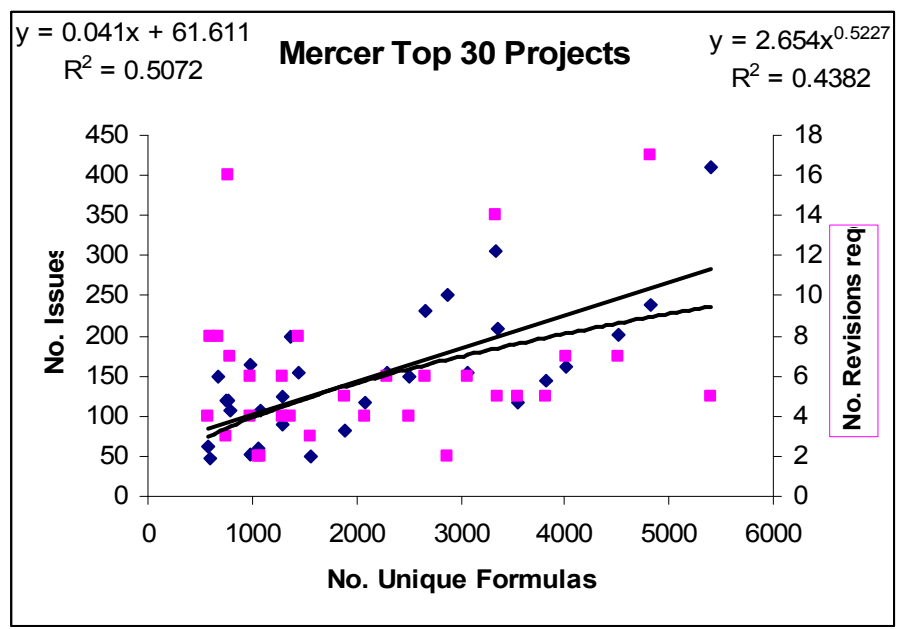

Figure 1: Scatterchart of issues vs no. UF for 30 models reviewed by Mercer.

7% is low by most end-user standards; according to Panko [4] and repeated by many other studies, 20% is a more normal end-user defect rate. It would be reasonable to infer that those developing the models had already done their own checking and cleanup before handing these models over for external audit. Later, we argue that defect removal can be made more efficient by reviewing earlier in the process, rather than waiting to review the final product to find them. Table 1 shows how many revisions one would expect for a given rate of defect injection and removal.

| D.I.R.%→ D.R.E%↓ | 3% | 4% | 5% | 7% | 10% | 15% | 20% | 30% |
|---|---|---|---|---|---|---|---|---|
| 20% | 15 | 17 | 18 | 20 | 22 | 25 | 29 | 37 |
| 25% | 12 | 13 | 14 | 16 | 18 | 20 | 23 | 30 |
| 30% | 12 | 12 | 13 | 15 | 16 | 18 | 20 | 26 |
| 35% | 10 | 11 | 11 | 12 | 14 | 16 | 18 | 23 |
| 40% | 9 | 9 | 10 | 11 | 12 | 14 | 15 | 20 |
| 50% | 6 | 7 | 7 | 8 | 9 | 10 | 12 | 16 |
| 60% | 7 | 7 | 7 | 8 | 9 | 10 | 11 | 14 |
| 80% | 5 | 5 | 5 | 5 | 6 | 7 | 7 | 10 |
| 100% | 3 | 3 | 3 | 4 | 4 | 5 | 6 | 8 |

Table 1: Estimated no. of revisions for 2000 Units of work (Unique Formulas) For given Defect Removal Efficiency and Defect Injection Rates

Even at only 3% injection rate and 100% removal efficiency you still need **three** revisions:

1 to write the initial model and inject 30 defects;
2 to remove the 30 defects by making 30 changes which causes 1 more defect;
3 to remove the final(?) defect.

Surveys of end-user development that review the products at an earlier stage find typically 20% of defects. Informal inspection efficiency is around 50%; it takes a formal process to reach 75%. For an average size model of 2000 UFs, 20% DIR and 50% DRE, one would expect 12 revisions to reach a sign-off point.

We argue that the payback for this investment in early review time is justified by the saving in project delay and expensive errors in use. So, how can people be encouraged to review their work earlier? We believe that this can be done by helping them to become aware of the real quality of their work.

## Overconfidence

Justin Kruger and David Dunning of Cornell University reported [2] that

*'People tend to hold overly favorable views of their abilities ... their incompetence robs them of the metacognitive ability to realize it. Paradoxically, improving the skills of participants, and thus increasing their metacognitive competence, helped them recognize the limitations of their abilities.'*

*'Because people usually choose what they think is the most reasonable and optimal option ... the failure to recognize that one has performed poorly will instead leave one to assume that one has performed well.'*

*'Sullivan, in 1953, marveled at "the failure of learning which has left their capacity for fantastic, self-centered delusions so utterly unaffected by a life-long history of educative events".'*

Ray Panko [4] has presented analysis to Eusprig on the rates of errors.

*'Consistent with human error research, which has seen similar error rates across cognitive tasks of comparable difficulty, spreadsheet error rates are very similar to those in traditional programming. However, while programmers spend 25% to 60% of their development time in testing, testing among spreadsheet developers in industry is extremely rare.'*

## Making work quality visible

The key, then, is to work in a way that faces the facts of defect injection, that rewards the removal of defects rather than punishes the necessary precondition of discovery. The management of developers is not to 'hold their feet to the fire' but to make the facts inescapably obvious. It is easy to brush aside and forget errors that have been fixed or can be quickly fixed. A written record helps avoid overconfidence and provides a basis for post-hoc analysis that can point to process improvements that can be made.

James Reason [3] advises self-knowledge as the first line of defence against error. His research in the medical world has found that the best consultants are those who consistently analyze their work and learn from their mistakes, rather than the characteristic arrogant denial that 'we don't make mistakes here'.

To assist individuals to become conscious of their own quality, some minimal schemes have been devised to collect and analyse data. One of these which can be paper-based or supported by (surprise!) a spreadsheet is the Personal Software Process (PSP) described below.

### Early visibility

The conventional software development lifecycle (SDLC) is described in phases of Requirements Analysis, System Design, Software Development, and Testing. Where reviews take place during this lifecycle, project managers can use early defect data to estimate the 'troublesomeness' of the product being developed, and get an indication of when it is safe to release.
Given that you collect defect arrival data, and the curve has achieved its maximum at time $t_m$ (e.g., the inflection point), you can calculate, assuming a Rayleigh distribution [11], the likely final number of defects and when they should be removed. The simplest initial assumption is that ~40% of the total defects have appeared by $t_m$.

The identification of separate activities with skills appropriate to each fits the specialised world of the professional software developer constructing projects for delivery to users. For end-user development, the fact that the customer is also the developer abbreviates the early stages as the spreadsheet creator believes that they understand their own requirements. They are often also their own tester. However, when they are building spreadsheets for use by others, they are already in a development role and the enterprise should equally consider the value of a separate testing role.

For applications that are critical to the success of the business, it is already recognised that there is a benefit from separating these stages and consciously documenting requirements, designing a layout and a data flow, and an independent review from a second pair of eyes. Panko [4] has related the types of testing appropriate to the stage of development.
At Eusprig 2006, Kumiega and Van Vliet of the Illinois Institute of Technology [5] described a spiral development methodology for rapid prototyping in financial markets. By focusing on return on investment, they increase the amount of amount of review and testing that a spreadsheet trading model gets in line with the value at risk in using the model.

### Disciplined Development

The Personal Software Process [6] of Watts Humphrey [7] is a disciplined approach to improving one's software development process. Through a series of cumulative exercises, developers learn:

1. Time and Defect recording
2. Software Size measurement
3. Software Size estimation
4. Statistically based estimation using Proxies
5. Time estimation and project scheduling
6. Process management
7. Design and Code reviews
8. Quality management through defect reduction

9. Design notation, techniques, verification
10. Scaling up the PSP to larger projects
11. How to develop the PSP into the future

## Paradoxes of measurement

Normal Fenton [9] reported on a conversation with the Robert Grady of HP about a system with no reported field defects. Initially, it was thought to be an example of 'zero defects'. They later discovered that was because no one was using it.

Capers Jones [10] pointed out in 1991 that there are two general rules for customer-reported defects:

- The number of defects found correlates directly with the number of users
- The number of defects found correlates inversely with the number of defects that exist

This is because if the software has many users, it will have more execution time, and hence, more defects will be uncovered more quickly. Conversely, if the software is buggy, people will not use it, and fewer defects will be uncovered!

## Answers

1) Are defects good or bad?

This question is using emotionally loaded words. If defects are 'bad', then those who create them are 'bad' and those who report them are messengers of 'bad' news. People will hide defects or disclaim them or divert responsibility for them. Defects exist. Would you rather not know they are there?

2) If your tester tells that they found no defects in a piece of software, is that good or bad?

It depends on what else you know from previous experience of the software developer's quality record, the efficiency of the tester, and any previous history of problems with the application under test.

## Conclusion

If risk to the enterprise is to be reduced, individuals need to become aware of the reality of the quality of their work. We recommend that where there is no independent review, individuals should adopt that role and consciously examine their work and record issues for correction as if they were doing it for someone else[12]. Such a record will serve as an inescapable reminder of the actual difficulty and quality of the project they are working on, and make it less likely that the product will be released in ignorance of the facts about its state. Based on previous research [8] we propose that in order to inculcate such a discipline research experiments are required towards a practical implementation of a Personal Spreadsheet Process.